\documentclass[12pt]{article}
\usepackage{amsmath,amssymb}
\usepackage{cite}
\usepackage{wrapft}
\newcommand{\be}{\begin{equation}}
\newcommand{\ee}{\end{equation}}
\newcommand{\bea}{\begin{eqnarray}}
\newcommand{\eea}{\end{eqnarray}}

\begin{document}
\vskip1cm
\begin{flushright}
\end{flushright}
\vskip1cm
\begin{center}
{\Large\bf Noncommutativity  and Ramond-Ramond
fields\footnote{Published in Phys. Lett. B 553 (2003) 301-308.}}

\vskip 1cm {\bf{{Masud Chaichian, Dimitri Polyakov and Anca
Tureanu}}

{\it High Energy Physics Division, Department of Physical Sciences,
University of Helsinki\\
\ \ {and}\\
\ \ Helsinki Institute of Physics,\\ P.O. Box 64, FIN-00014
Helsinki, Finland}}
\end{center}
\vskip1cm
\begin{abstract}

We study the properties of the Ramond fields in superstring theory
in the presence of a constant B-field and the subsequent
implications of space-time noncommutativity for space-time fermions
and space-time supersymmetry. We find that the noncommutativity of
space-time coordinates leads to extra singularities in the spin
field correlators and to appearance of the extra poles in the RR
scattering amplitudes, carrying the opposite GSO parities.
\end{abstract}

\vskip1cm
\section{Introduction}

It is well known that the dynamics of open strings in a constant
$B_{mn}$-field background implies the noncommutativity of space-time
coordinates \cite{shahin,seiberg}. At the same time, the influence
of the B-field on the dynamics of the Ramond fields and physical
consequences of the space-time noncommutativity for fermionic fields
are not yet well understood. One particularly important question is
how the scattering amplitudes of the Ramond fields are affected by
the presence of the B-field background and what is their
modification for the noncommutative space-time.
 In the bosonic case the modification is trivial:
the correlators of the NS and NS-NS fields are unchanged, up to a
constant "noncommutative phase" \cite{gukov}. For the Ramond sector,
however, the situation is different. The bosonization rules for the
worldsheet fermions are modified in the B-field background and this
affects the structure of the space-time spin operators and their
mutual operator products. As a result, the correlators involving the
Ramond and Ramond-Ramond fields receive nontrivial corrections in
the noncommutative $\theta$ parameter.

In this letter we analyze the example of the gravitational lensing
of the RR-fields in noncommutative space, involving the two-point RR
scattering amplitudes on a disc in the B-field background. We derive
the modified bosonization relations for the spin fields in the
noncommutative background and compute the relevant scattering
amplitudes of the RR fields. We find that the change in bosonization
leads to the nontrivial $\theta$ corrections to these amplitudes. In
particular, the correction terms have a modified pole structure;
namely, in the first order in the noncommutative $\theta$ parameter
the correlator will be shown to receive contributions from the
intermediate states of an opposite (odd) GSO sector. Thus an
important physical consequence for the spin fields in the B-field
background is that the noncommutativity of the space-time
effectively mixes the operators from the opposite GSO sectors
\cite{gso} if these operators involve the space-time spin fields.

\section{Noncommutativity and bosonization of fermions}

Let us recall how the noncommutativity of space-time appears in
string theory. Firstly, the imposition of the B-field changes the
worldsheet boundary conditions for an open string. For a string
moving in a constant B-field background these boundary conditions
are given by:
\be\label{1}g_{mn}(\partial-{\bar\partial})X^n+2\pi{\alpha^\prime}
B_{mn}(\partial+{\bar\partial})X^n|_{z=\bar{z}}=0.\ee
Here ${g_{mn}}$ is the space-time metric and it is implied that the
worldsheet surface is conformally mapped to the half-plane. The
boundary conditions (1) interpolate between Dirichlet and Neumann
conditions. If the B-matrix is invertible, the limit of
$B\rightarrow{\infty}$ corresponds to Dirichlet boundary conditions
for the spatial directions in which the B-field is applied
\cite{seiberg}. For this reason the imposition of an infinite
B-field is similar to the T-duality transformation in the
appropriate directions. The propagator corresponding to the boundary
conditions (\ref{1}) is given by \cite{seiberg}
\bea\label{2}&&\langle X^m(z,\bar{z})X^n(w,\bar{w})\rangle\\
&=&
-{\alpha^\prime}\left[g^{mn}\log|z-w|-g^{mn}\log|z-{\bar{w}}|+G^{mn}\log|z-{\bar{w}}|^2+{1\over{2\pi{\alpha^\prime}}}\theta^{mn}
\log\left(-{{z-{\bar{w}}}\over{{\bar{z}}-w}}\right)\right],
\nonumber\eea
where
\bea\label{3} G^{mn}&=&\left[(g+2\pi{\alpha^\prime}B)^{-1}g
(g-2\pi{\alpha^\prime}B)^{-1}\right]^{mn}\cr
\theta^{mn}&=&-2\left[(\pi{\alpha^\prime})^2
(g+2\pi{\alpha^\prime}B)^{-1}B
(g-2\pi{\alpha^\prime}B)^{-1}\right]^{mn}. \eea
In the limit when both $z$ and $w$ approach the real axis:
$z={\bar{z}}\rightarrow{t_1}$, $w={\bar{w}}\rightarrow{t_2},$ the
propagator (2) becomes
\be\label{4} \langle
X^m(t_1)X^n(t_2)\rangle=-{\alpha^\prime}G^{mn}\log(t_1-t_2)^2
+{i\over2}\theta^{mn}\mbox{sign}(t_1-t_2),\ee
implying the commutation relation:
\be\label{5} [X^m,X^n]=i\theta^{mn} \ee
after the regularization and the time ordering. The change in  the
propagator (\ref{4}) and the commutation relation (\ref{5}) give
rise to the appearance of the noncommutative "star product" in the
field-theoretic
 low energy effective action as $\alpha^\prime\rightarrow{0}$.
Thus the space-time noncommutativity is  related to the
 antisymmetric term in the propagator and
the $\theta^{mn}$ matrix defines the noncommutativity parameter.

The important question is what are the implications of the
noncommutativity for the superspace and what are the properties of
fermions in noncommutative space-time. Addressing this question from
the string theory point of view means that one has to study the
properties of the Ramond fields in the $B$-field background. For the
NSR worldsheet fermions in the $B$-field background the boundary
conditions, being the supersymmetric extension of (\ref{1}), have
the form:
\be\label{6} g_{mn}(\psi^n-{\bar\psi}^n)+2\pi{\alpha^\prime}
B_{mn}(\psi^n+{\bar\psi}^n)=0.\ee
The propagators corresponding to these boundary conditions are given
by
\bea\label{7} \langle\psi^m(z,{\bar{z}})\psi^n(w,{\bar{w}})\rangle
&=&\left(G^{mn}-{1\over2}g^{mn}\right)\left({1\over{z-{\bar{w}}}}+{1\over{{\bar{z}}-w}}\right)
\cr&+&{1\over2}g^{mn}\left({1\over{z-w}}+{1\over{{\bar{z}}-{\bar{w}}}}\right)
+{1\over{2\pi{\alpha^\prime}}}\theta^{mn}
\left({1\over{z-{\bar{w}}}}-{1\over{{\bar{z}}-w}}\right). \eea
 The propagator
(\ref{7}) implies that one can continue $\psi(z,{\bar{z}})$ to the
entire complex plane with the corresponding correlators between
$\psi(z)$ and ${\bar\psi}({\bar{z}})$ expressed as
\bea\label(8) \langle\psi^m(z)\psi^n(w)\rangle
&=&{1\over2}{{g^{mn}}\over{z-w}},\cr
\langle{\bar\psi}^m({\bar{z}}){\bar\psi}^n({\bar{w}})\rangle&=&{1\over2}
{{g^{mn}}\over{{\bar{z}}-{\bar{w}}}},\cr
\langle{\bar\psi}^m({\bar{z}}){\psi}^n({{w}})\rangle
&=&{{G^{mn}-{1\over2}g^{mn}+{1\over{2\pi{\alpha^\prime}}}
\theta^{mn}}\over{{\bar{z}}-{{w}}}},\cr
\langle{\psi}^m({{z}}){\bar\psi}^n({\bar{w}})\rangle
&=&{{G^{mn}-{1\over2}g^{mn}-{1\over{2\pi{\alpha^\prime}}}
\theta^{mn}}\over{{{z}}-{\bar{w}}}}.\eea
The important property of these propagators
is the presence of the non-diagonal interactions  $<\psi^m\bar\psi_n>$
and  $<\bar\psi^m\psi_n>$ for $n\neq{0}$.
This fact will prove to be of importance for the
properties of the space-time fermions in noncommutative space.

Now we are prepared to construct the Ramond spin fields in the presence
 of the B-field background.
Of course our discussion is entirely restricted to the case of the
space-type (magnetic) noncommutativity
\be\label{9} \theta_{ij}\neq{0},\ \  i,j{\neq}0, \ee
since the
formulation of string theory in backgrounds with time-like
noncommutativity is problematic \cite{seib,gomis}. The construction
is quite analogous to the standard scheme \cite{fms}. As usual, one
starts with constructing five complex worldsheet fermions out of ten
$\psi^m$'s:
\bea\label{10}\Lambda_1&=&i\psi_0+i\psi_9, \Lambda_1^{*}=i
\psi_0-i\psi_9;\cr
\Lambda_2&=&\psi_1+i\psi_2, \Lambda_2^{*}=\psi_1-i\psi_2; \cr
\Lambda_3&=&\psi_3+i\psi_4, \Lambda_3^{*}=\psi_3-i\psi_4; \cr
\Lambda_4&=&\psi_5+i\psi_6, \Lambda_4^{*}=\psi_5-i\psi_6; \cr
\Lambda_5&=&\psi_7+i\psi_8, \Lambda_5^{*}=\psi_7-i\psi_8; \eea
and analogously for the antiholomorphic part.
Next, one bosonizes $\Lambda_j$ as
\be\label{11}\Lambda_j=e^{i\varphi_j},\ \ j=1,...,5.\ee
Finally, one constructs the space-time spin operator
$\Sigma_\alpha,\alpha=1,...,32$ as
\be\label{12}\Sigma_\alpha(z)={\prod_{j=1}^5}e^{{i\over2}a_j\varphi_j},\ee
with
\be\label{13}a_j=\pm{1}\ee
 and each value of the spinor
index $\alpha$ corresponding to some particular combination of
$\lbrace{a_j}\rbrace,j=1,...,5.$ In the absence of the B-field this
spin operator has a conformal dimension $5/8$ as its OPE with itself
is given by:
\be\label{14}\Sigma_\alpha(z)\Sigma_\beta(w)\sim
{{\delta_{\alpha\beta}}\over{(z-w)^{5\over4}}}+\sum_p
{{{\gamma^{m_1...m_p}_{\alpha\beta}}\psi_{m_1}...\psi_{m_p}}
\over{(z-w)^{{5\over4}-{p\over2}}}}+\mbox{derivatives}.\ee
Similarly, one can write the OPE between $\Sigma_\alpha(z)$ and
${\bar\Sigma}_\alpha({\bar{w}})$ on a disc using the relation:
\be\label{15} {\bar\Sigma}_\alpha({\bar{w}})=
{\gamma^{0m_1...m_q}_{\alpha\beta}}\Sigma_\beta({\bar{w}}), \ee
where $q$ is the number of Dirichlet boundary conditions on the disc
(out of the total number of 10). Let us now turn to the case of the
nonzero B-field. In the presence of the B-field the OPE involving
 the matter spin operators with different chiralities,
change significantly. First of all one needs to point out the
relation between left and right modes of spin operators on the disc
in the noncommutative case. For simplicity, we consider the case
when all the boundary conditions are Neumann. First, using the
propagators (8) it is easy to see that the relation between $\psi_m$
and $\bar\psi_m$ is given by:
\be\label{16}\bar\psi_{n}(\bar{z})=\left(G_{mn}-{1\over2}g_{mn}\right)\psi_{m}(\bar{z})+
\theta_{mn}\psi_{m}(\bar{z}). \ee
 For simplicity, let us consider for
the time being the special case of $\theta=\theta_{13}\neq{0}$,
while setting all other components to zero; the result will be
easily generalized to the arbitrary case.

In this case the corresponding relations between  $\Lambda$
and $\bar\Lambda$, following from (16), are given by
\bea\label{17}{\bar\Lambda}_1(\bar{z})&=&\Lambda_1(\bar{z})+
\theta_{13}\Lambda_2(\bar{z}),\ \ \
{\bar\Lambda}_2(\bar{z})=\Lambda_2(\bar{z})-
\theta_{13}\Lambda_1(\bar{z}),\cr
{\bar\Lambda}_a(\bar{z})&=&\Lambda_a(\bar{z}),\ \ a=3,4,5.\eea
One can use these relations to find the correspondence between
$\Sigma$ and $\bar\Sigma$ on the disc, provided that $\theta$ is
small. Indeed, using $:{\sqrt{x+\epsilon}}:\sim:{\sqrt{x}}:
+:{{\epsilon}\over{2{\sqrt{x}}}}:+O(\epsilon^2)$ when $\epsilon$ and
$x$ operators are independent, one has
\bea\label{18}
e^{{i\over2}{\bar\varphi_1}}(\bar{z})&=&(\Lambda_1(\bar{z})-\theta_{13}\Lambda_2(\bar{z}))^{1\over2}
=e^{{i\over2}\varphi_1}(\bar{z})-{{\theta_{13}}\over2}
e^{-{i\over2}\varphi_1+{i}\varphi_2}+O(\theta^2)\cr
&=&e^{{i\over2}\varphi_1}(\bar{z})-{{\theta_{13}}\over2}
:\Lambda_2{e^{-{i\over2}\varphi_1}}:(\bar{z})+O(\theta^2),\cr
e^{{i\over2}{\bar\varphi_2}}
(\bar{z})&=&(\Lambda_2(\bar{z})+\theta_{13}\Lambda_1(\bar{z}))^{1\over2}
=e^{{i\over2}\varphi_2}(\bar{z})+{{\theta_{13}}\over2}
e^{-{i\over2}\varphi_2+{i}\varphi_1}+O(\theta^2)\cr
&=&e^{{i\over2}\varphi_1}(\bar{z})+{{\theta_{13}}\over2}
:\Lambda_1{e^{-{i\over2}\varphi_1}}:(\bar{z})+O(\theta^2),\cr
e^{{i\over2}{\bar\varphi_a}} (\bar{z})&=&e^{{i\over2}{\varphi_a}}
(\bar{z}).\eea
Using these relations along with the definition (12) for the spin
operators, it is straightforward to show that
$$\bar\Sigma_\alpha(\bar{z})=
\Sigma_\alpha(\bar{z})+{i\over2}\theta_{13}((\gamma^1)_{\alpha\beta}\psi^3
-(\gamma^3)_{\alpha\beta}\psi^1)
\Sigma_\beta(\bar{z})+O(\theta^2).$$
This relation is straightforward to generalize to the
case of arbitrary $\theta_{ij}$:
\be\label{19}\bar\Sigma_\alpha(\bar{z})=
\Sigma_\alpha(\bar{z})+{i\over2}\theta_{ij}((\gamma^i)_{\alpha\beta}\psi^j
-(\gamma^j)_{\alpha\beta}\psi^i)
\Sigma_\beta(\bar{z})+O(\theta^2).\ee
This important relation leads to some remarkable physical consequences.
Upon multiplying $\Sigma$
 by the ghost spin operator ${e^{-{1\over2}\phi}}$,
it is easy to see that the full matter $+$ ghost
antiholomorphic  spin operator is expressed in terms of the
left-moving modes with mixed GSO parities; indeed,
it is easy to check that the operators of the type
${\sim}e^{-{1\over2}\phi}\Sigma\psi$ are GSO-odd.
Therefore it shows that the noncommutativity mixes the
even and odd GSO sectors of the Ramond operators in the first
order of $\theta$.
Of course such a mixing does not reveal itself in the sphere amplitudes
but becomes  important on the disc when the left and the right modes
mix with each other. In particular,
 this leads to the appearance of intermediate
poles of opposite GSO parity in the two-point disc amplitude
of the RR fields in noncommutative backgrounds.

\section{RR scattering amplitudes in the B-field}

Consider the two-point scattering amplitude of RR vertex operators
on the disc in the case of  $B_{ij}\neq{0},\theta_{ij}\neq{0}.$ The
RR vertex operators taken at the standard $-1/2,-1/2$-picture are
given by
\be\label{20}
V_{RR}(z,\bar{z})=e^{-{1\over2}\phi-{1\over2}{\bar\phi}}
\Sigma_\alpha{\bar\Sigma}_\beta{e^{ikX}}(z,\bar{z})
\gamma^{m_1...m_q}_{\alpha\beta}F_{m_1...m_q}(k).\ee

For the two-point disc amplitude one can take both RR vertices at
this canonical picture.
Using
 the bosonization formulae for spin operators (\ref{12}) along with the relation
(\ref{19}) it is straightforward to compute the amplitude. First of
all, the four-point correlator of matter spin operators (giving rise
to the two-point disc amplitude) is given by:
\bea\label{21} &&\langle\Sigma_{\alpha_1}(z)
{\bar\Sigma}_{\alpha_2}(\bar{z}) \Sigma_{\beta_1}(w)
{\bar\Sigma}_{\beta_2}(\bar{w})\rangle\\
&&= -{{\gamma^\mu_{\alpha_1\beta_1}(\gamma_\mu)_{\alpha_2\lambda}
(z-\bar{z})^{1\over4}(w-\bar{w})^{1\over4}}\over
{(z-w)^{3\over4}(z-\bar{w})^{3\over4}(\bar{z}-w)^{3\over4}
(\bar{z}-\bar{w})^{3\over4}}} \left(\delta_{\lambda\beta_2}
+{1\over2}{\gamma^{ij}_{\lambda\beta_2}}\theta_{ij}
\left({{\sqrt{w-\bar{w}}}\over{z-\bar{z}}}
+{{\sqrt{z-\bar{z}}}\over{w-\bar{w}}}\right)\right) \cr
&&+{{\gamma^\mu_{\alpha_1\alpha_2}(\gamma_\mu)_{\beta_1\lambda}
(z-\bar{w})^{1\over4}(\bar{z}-{w})^{1\over4}}\over
{(z-w)^{3\over4}(z-\bar{z})^{3\over4}(w-\bar{w})^{3\over4}
(\bar{z}-\bar{w})^{3\over4}}}
\left(\delta_{\beta_2\lambda}-{1\over2}{\gamma^{ij}_{\lambda\beta_2}}\theta_{ij}
\left({{\sqrt{z-\bar{w}}}\over{{\bar{z}-w}}}
+{{\sqrt{{\bar{z}}-{w}}}\over{{{z}-\bar{w}}}}\right)\right)
+O(\theta^2).\nonumber\eea

In comparison with the $B=\theta=0$ case, there are
the extra square root factors in the first and in the second terms,
appearing in the first order in $\theta$.
Using this correlator and noting that
the B-field does not change the superconformal ghost propagator,
one easily finds that the two-point Ramond-Ramond half-plane
correlation function
in the noncommutative case is given by:
\bea\label{22}&&\langle{V_{RR}(z,\bar{z};k)}{V_{RR}(w,\bar{w};p)}\rangle\\
&=&\Big(-\gamma^\mu_{\alpha_1\beta_1}(\gamma_\mu)_{\alpha_2\lambda}
|z-\bar{z}|^{\lbrace{k,k}\rbrace} |w-\bar{w}|^{\lbrace{p,p}\rbrace}
|z-\bar{w}|^{2{\lbrace}k,p\rbrace-2} |z-{w}|^{2(kp)-2}\cr
&{\times}&\left(\delta_{\beta_2\lambda}+{1\over2}\gamma^{ij}_{\beta_2\lambda}
\theta_{ij} \left({{\sqrt{w-\bar{w}}}\over{z-\bar{z}}}
+{{\sqrt{z-\bar{z}}}\over{w-\bar{w}}}\right)\right) \cr
&+&{\gamma^\mu_{\alpha_1\alpha_2}}(\gamma_\mu)_{\beta_1\lambda}
|z-\bar{z}|^{\lbrace{k,k}\rbrace-1}
|w-\bar{w}|^{\lbrace{p,p}\rbrace-1}
|z-\bar{w}|^{2{\lbrace}k,p\rbrace} |z-{w}|^{2(kp)-2}\cr
&\times&
\left(\delta_{\beta_2\lambda}-{1\over2}\gamma^{ij}_{\beta_2\lambda}\theta_{ij}
\left({{\sqrt{z-\bar{w}}}\over{{\bar{z}-w}}}
+{{\sqrt{{\bar{z}}-{w}}}\over{{{z}-\bar{w}}}}\right)\right)\Big)
\times \gamma^{m_1...m_p}_{\alpha_1\alpha_2}
\gamma^{n_1...n_p}_{\beta_1\beta_2} F_{m_1...m_q}(k)F_{n_1...n_q}(p)
.\nonumber\eea
Here we denoted
\be\label{23}
\lbrace{a,b}\rbrace\equiv{\alpha^\prime}\left(2G^{mn}-g^{mn}
+{{\theta^{mn}}\over{2\pi\alpha^\prime}}\right)a_mb_n,\ \ \
(ab)=\alpha^\prime{a_m}b^m\ee
for any space-time vectors $a$ and $b$. For simplicity, let us
consider the rank $q$ even, i.e. the type IIB case. Then the second
term linear in $\theta$ in (22) is absent as its gamma-matrix trace
factor vanishes. The correlator (22) now needs to be integrated over
the worldsheet to obtain the amplitude. The zero order in $\theta$
part of the correlator is the standard one, giving rise to the usual
commutative gravitational lensing \cite{klebanov}. From now on, let
us therefore concentrate just on the term linear in $\theta_{ij}$,
proportional to ${\sqrt{z-{\bar{z}}}}$, which is the one
contributing to the GSO parity mixing. To integrate it over the
worldsheet with the disc topology, it is convenient to conformally
map the half-plane to the disc using the transformation
$(z,\bar{z})\rightarrow(u,\bar{u})$ with
\be \label{24}z={i\over2}{{u+i}\over{u-i}}\ee
with $(u,\bar{u})$ now being the disc coordinates. To calculate the
scattering amplitude it is convenient to place one of the vertex
operators at the origin of the disc (corresponding to the point
$w=-{i\over2}$ on the halfplane) and to integrate over the position
of  another one. Writing $u=re^{i\lambda}$ and using
\bea\label{25} |z-w|^2={{r^2}\over{r^2+1-2r{\sin}\lambda}},\ \
|z-\bar{w}|^2={{1}\over{r^2+1-2r{\sin}\lambda}},\cr
z-\bar{z}={i{(r^2-1)}\over{r^2+1-2r{\sin}\lambda}},\ \
w-\bar{w}=-i,\eea
we have:
\bea\label{26} A(k,p)&=&\langle V_{RR}(k)V_{RR}(p)\rangle|_{NC}\\
&=&{1\over2}\theta_{ij} Tr(\gamma^\mu\gamma^{n_1...n_q}
\gamma_\mu\gamma^{m_1...m_q}\gamma^{ij})F_{m_1...m_q}(k)F_{n_1...n_q}(p)
\cr
&\times&\int_0^{2\pi}d\lambda\int_0^\infty{dr}r
(r^2+1-2r{\sin}\lambda)^{-\lbrace{k,k}\rbrace
-\lbrace{k,p}\rbrace-(kp)+{3\over2}}
(r^2-1)^{\lbrace{k,k}\rbrace+{1\over2}} r^{2(kp)-2}.\nonumber\eea
First of all, let us calculate the angular integral over $\lambda$.
We have:
\be\label{27}\int_0^{2\pi}
(r^2+1-2r{\sin}\lambda)^{-\lbrace{k,k}\rbrace
-\lbrace{k,p}\rbrace-(kp)+{3\over2}}\equiv \int_0^{2\pi}
(r^2+1-2r{\sin}\lambda)^\rho
=2\Gamma\left({1\over2}\right)F(\rho,\rho,1,r^2).\ee
where $F$ is hypergeometric function and we denoted
$\rho=-\lbrace{k,k}\rbrace -\lbrace{k,p}\rbrace-(kp)+{3\over2}$. The
amplitude is then reduced to the radial integral
\bea\label{28} A(k,p)&{\sim}&2\Gamma({1\over2})\int_0^1{dr}r
(r^2-1)^{\lbrace{k,k}\rbrace-{1\over2}}
r^{2(kp)-2}F(\rho,\rho,1,r^2)
\cr
&=&\Gamma({1\over2})\int_0^1{dx}(x-1)^{\lbrace{k,k}\rbrace+{1\over2}}
x^{(kp)-1}F(\rho,\rho,1,x), \eea
where we have introduced the new integration variable $x=r^2$. In
addition, here and everywhere below we suppress the gamma-matrix
trace for the sake of shortness. The evaluation of the radial
integral over $x$ involving  the hypergeometric function depends on
the value of the parameters $(kp)$ and $\lbrace{k,k}\rbrace$. In the
case of
\be\label{29} (kp)>0; \lbrace{k,k}\rbrace > -{1\over2}\ee
the evaluation of the  integral gives
\be\label{30} A(k,p)\sim
{{\Gamma({1\over2})\Gamma((kp))\Gamma(\lbrace{k,k}\rbrace+{1\over2})}
\over{\Gamma((kp)+\lbrace{k,k}\rbrace)}}
{{}_3}F_2(\rho;\rho;(kp);1;(kp)+\lbrace{k,k}\rbrace+{3\over2};1),
\ee
where ${{}_3}F_2(\rho;\rho;(kp);1;
(kp)+\lbrace{k,k}\rbrace+{3\over2};1)$ is the generalized
hypergeometric function. The constraints (\ref{29}) insure that the
amplitude has no physical poles in this case. When the conditions
(\ref{29}) are not fulfilled, the value of the integral (\ref{28})
is different. Expanding $F(\rho,\rho,1,x)$ in series we can write
this integral as
\bea\label{31}&&A(k,p)\sim
\Gamma\left({1\over2}\right)\int_0^1{dx}x^{(kp)-1}(x-1)^{\lbrace{k,k}\rbrace+{1\over2}}
\left(1+\sum_{n=1}^{\infty}{{\rho(\rho+1)...(\rho+n-1)}\over{n!}} x^n\right)\\
&=&\Gamma\left({1\over2}\right)\left\{B\left((kp),{\lbrace{k,k}\rbrace+{1\over2}}\right)
+\sum_{n=1}^{\infty}{{\rho(\rho+1)...(\rho+n-1)}\over{n!}}
B\left((kp)+n,{\lbrace{k,k}\rbrace+{3\over2}}\right)\right\}.\nonumber\eea
The crucial property of this amplitude is that it has extra physical
poles corresponding to
\be\label{32}\lbrace{k,k}\rbrace=-{3\over2}-m, \ee
where $m$ is non-negative integer. These poles correspond to the
appearance of the GSO-odd intermediate states. As it is easy to
notice, these extra poles, leading to the GSO-parity change,
originate from the square roots in the unintegrated amplitude,
creating  cuts on the worldsheet surface. In the commutative case
these poles are of course absent since $\lbrace{k,k}\rbrace=0$ when
$B=0$.

\section{Conclusions}

In this letter we have studied the properties of Ramond and
Ramond-Ramond fields in constant B-field backgrounds and the
influence of the string theory noncommutativity on the space-time
fermions and bispinors. We have found that properties of spin
operators are significantly changed in the noncommutative case.
Namely, we found that in noncommutative space-time the relation
between left and right modes of spin operators on the disc  involves
the mixing of the operators with opposite GSO parities. The effect
of the GSO parity change becomes more transparent when one considers
the Ramond-Ramond scattering amplitudes on the disc in the
noncommutative space-time, where the "anomalous" interaction between
left and right-moving fields brings about significant physical
consequences. In the two-point RR disc amplitude, describing the
"gravitational lensing" of the RR particles off the noncommutative
plane, we observe the appearance of physical poles corresponding to
intermediate GSO-odd states. Next, because of the extra gamma-matrix
factor of $\gamma^{ij}$, the first order $\theta$-correction appears
as if the boundary conditions in the given i-j directions are
changed from Neumann to Dirichlet, or the T-duality transform is
applied to these directions, imitating a scattering off a $D8$-brane
spanned in the directions transverse to i and j. This is consistent
with the fact that the noncommutative boundary conditions
effectively mix the Dirichlet and Neumann boundary conditions, and,
in the limit of a large B-field, introducing the noncommutativity is
similar to the T-duality transformation applied in the directions of
B. The effects described  in this paper may have several interesting
applications. Particularly interesting is the limit of the large
B-field. In this case the interactions of fermions with opposite
chiralities become dominant and the physical sector of the theory is
shifted into the GSO-odd sector. At the same time, the axionic term
becomes dominant  in the worldsheet NSR action. It is known
\cite{amp} that the worldsheet action with the axionic term
${\sim}B_{ij}\partial{X^i}{\bar\partial}{X^j}$ and with the graviton
absent is zigzag invariant, i.e. they may describe the confining QCD
string. In view of the above, one may speculate that the confining
phase of the QCD may be relevant to the dynamics of the GSO-odd
sector of the NSR superstring theory. Another important issue is the
role of the tachyon in noncommutative space. Because of the GSO
parity change, the tachyon becomes a physical GSO-projected state in
the $B\rightarrow{\infty}$ limit. Changes in conformal dimensions
may insure that noncommutative tachyons are still consistent with
the vacuum stability. We hope to address this and the related issues
in the future works.

\vskip1cm

\section*{Acknowledgements}

We are grateful to Claus Montonen, A.M. Polyakov and M.
Sheikh-Jabbari for useful comments and suggestions. The authors
acknowledge the financial support of the Academy of Finland under
the Project no. 54023

\end{document}